\documentclass{ismdproc-new}
\usepackage{graphicx}
\newcommand{\CASCADE}{{\sc Cascade}}
\newcommand{\PYTHIA}{{\sc Pythia}}
\newcommand{\HERWIG}{{\sc Herwig}}

\begin{document}
\title{Multiparton interactions in $ep$ scattering at HERA}


%

%

%
\author{{\slshape Hannes Jung}  for the H1 and ZEUS Collaborations\\[1ex]
DESY, Notketra{\ss}e 85, 22607 Hamburg, Germany }

\contribID{xy}  
\confID{yz}
\acronym{ISMD2010}
\doi            

\maketitle

\begin{abstract}
Measurements of jet production in the photoproduction region at HERA are presented, which are sensitive to the underlying event structure. It is argued that in the photoproduction region a significant amount of multi-parton interactions in the collinear factorisation ansatz are needed  to describe the measurement. A different approach for higher order parton radiation (CCFM parton showers) can also describe the measurement. 
\end{abstract}

\section{Introduction}
\label{sec:Introduction}
In proton-antiproton interactions, the underlying event structure has been investigated by measuring particle production in a region transverse to the direction of the jets. It was found that the measurements can be described only with Monte Carlo event generators (\PYTHIA ,\HERWIG) using
models of multi-parton interactions after elaborate
tuning~\cite{Affolder:2001xt, Moraes:2007rq,Khachatryan:2010pv}. 
Similar effects are
expected to occur also in the region of photoproduction in $ep$
collisions, where the photon is resolved into a hadronic component and
secondary scatterings can occur between the remnants of the proton and
the photon. 

Experimentally it is not possible on an event-by-event basis  to
separate effects coming from multi-parton interaction and those
coming from higher order radiation: a higher hadronic activity can
come from more and harder partons radiated in the initial state, from
effects of hadronization, and from hard multi-parton scattering. Special observables, like the average charged particle multiplicity, 
are used to search for effects in
addition to the primary hard scattering process. In order to get
confidence in the various models which the data are compared to, 
measurements are presented of multiplicities in both the hard and 
soft parts of the events. The regions which are dominated by the
hard interactions can be used as benchmark regions, where Monte Carlo
event generators supplemented with parton showers and hadronization
are expected to agree with measurements.

The measurements presented here are also of importance for the interpretation of LHC results in terms of multiparton interactions and underlying event structures.

\section{Models for the underlying event and multiparton interactions}
In collinear factorisation the partonic cross section $\sigma^{part}$ in lowest order is divergent, when the transverse momentum of the outgoing partons becomes small. A transverse momentum cut is usually applied to avoid this divergency. However, with high parton densities it can happen, if the transverse momentum cut is low, that the partonic cross section:
$\sigma^{part}(p_{t\;min}) = \int_{p_{t\;min}} d p_t \frac{d \sigma^{part}(p_t)}{dp_t}$
becomes larger
than the total cross section. 
To resolve this, the average number of interactions can be defined~\cite{Sjostrand:1987su,Sjostrand:2006za} according to
$\langle n \rangle = \frac{\sigma(p_{t\;min})^{part}}{\sigma_{tot}}$ and 
it is assumed that more than one partonic interaction can happen within
 one electron-proton
collision.
At HERA 
\begin{wrapfigure}{r}{0.47\textwidth}
\vskip -0.7cm
\centerline{\includegraphics[width=0.4\textwidth]{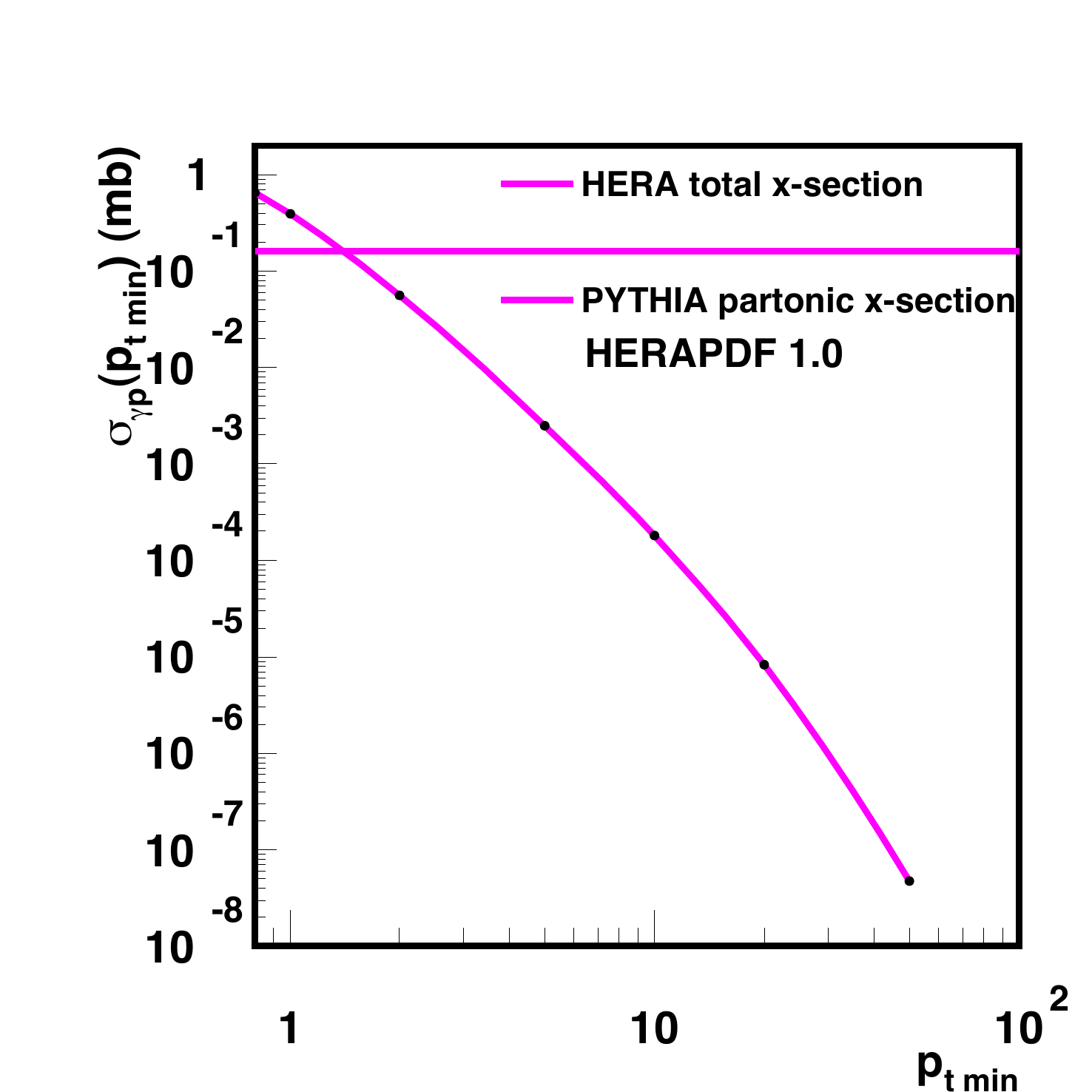}}
\caption{The $\gamma p$ cross section calculated with \PYTHIA\ using 
\protect\cite{:2009wt}. The vertical line indicates $\sigma_{tot}^{\gamma p}$ at $W_{\gamma p} = 200$ GeV.}
\label{Fig:ptmin}
\vskip -0.5cm
\end{wrapfigure}
energies ( $W_{\gamma p} = 200$ GeV), the partonic 
cross section calculated at LO in collinear factorisation becomes similar to the total photoproduction cross section at values of $p_t\sim 2$~GeV (Fig.~\ref{Fig:ptmin}). This value is small enough not to affect high $p_t$
jet cross section calculations but is also clearly above the soft
scale set by $\Lambda_{QCD} \approx 10^{-1}$~GeV. 
 The details of the
multiparton interaction model used in the
\PYTHIA~\cite{Sjostrand:1987su} Monte Carlo generator, which is used in the presented analyses, are described in
\cite{Sjostrand:1987su,Sjostrand:2006za}. 

However, additional radiation from the initial state can also be expected from a different parton shower mechanism as implemented in the \CASCADE\ Monte Carlo generator \cite{Jung:2010si}. In this approach, no cut-off for the partonic cross section is needed since the incoming partons are treated off-shell.

\section{The underlying event in photoproduction}
In the following two different analyses are discussed: inclusive jet production at large transverse momentum and a measurement of the underlying event in photoproduction.
\subsection{Inclusive jet production}
\begin{figure}[h]
\centerline{\includegraphics[width=0.5\textwidth]{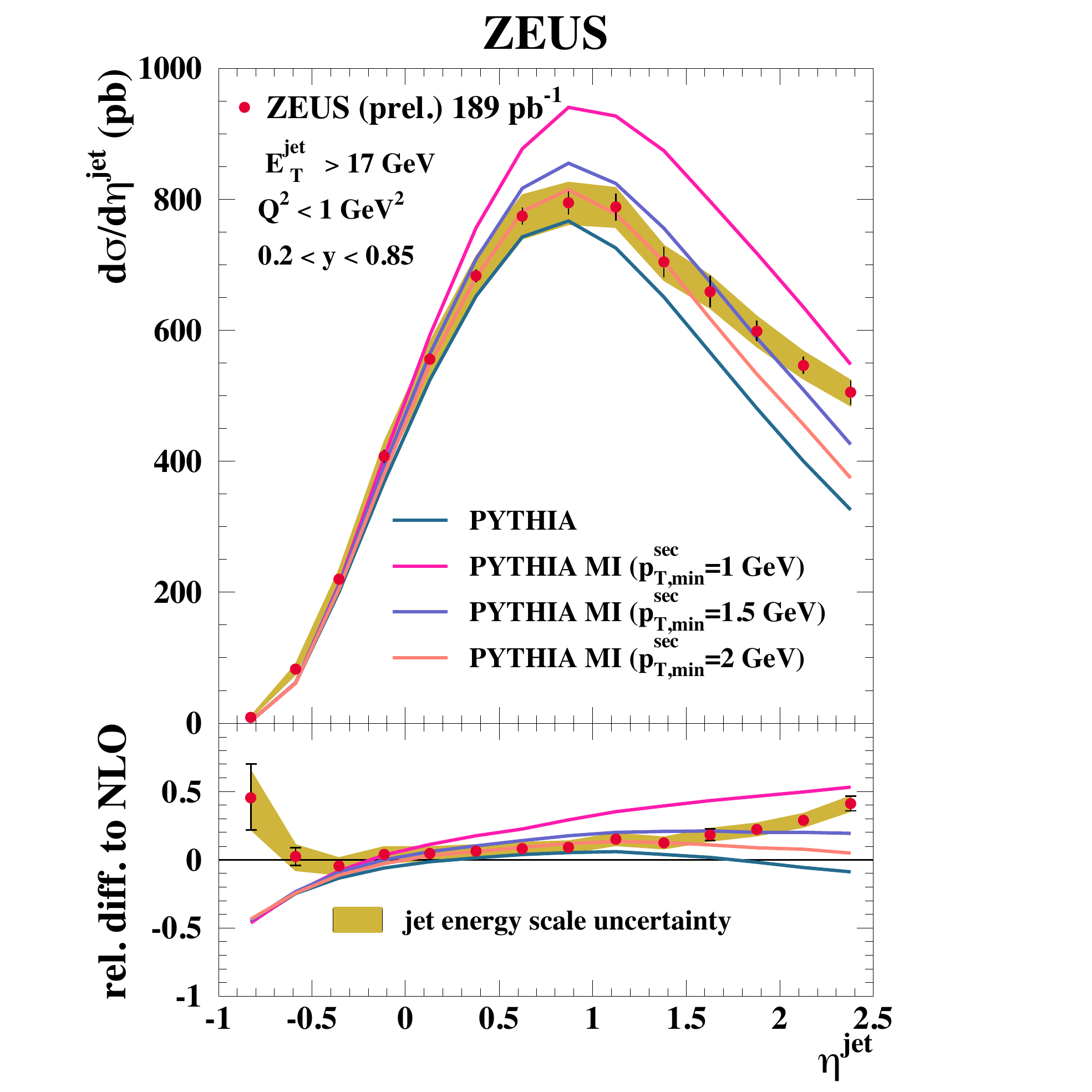}
\includegraphics[width=0.5\textwidth]{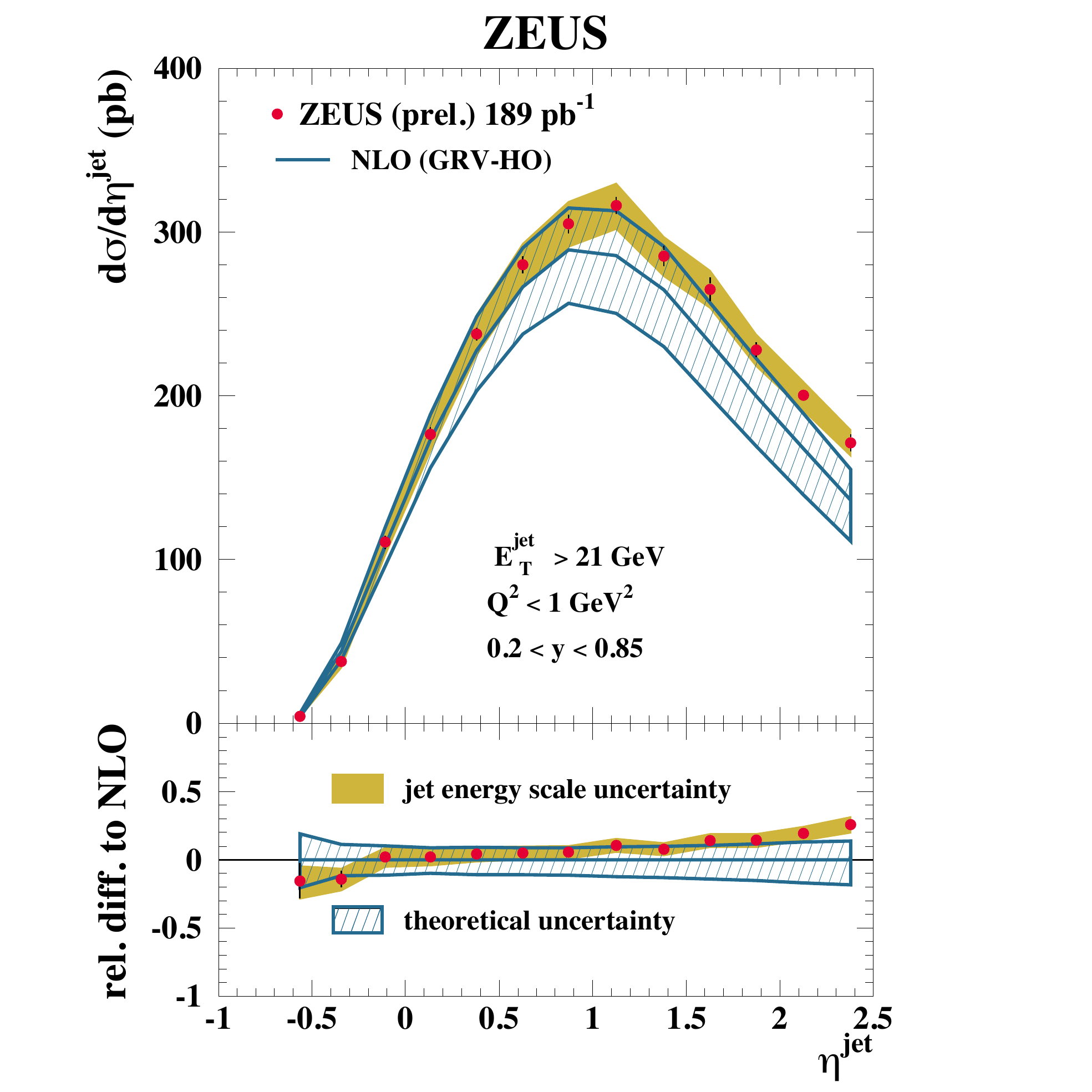}}
\caption{Differential cross section of inclusive jet in photoproduction for $E_T > 17$~GeV (left) and $E_T>21$~GeV (right). The data are compared with MC and NLO predictions.}\label{Fig:ZEUS}
\end{figure}
Inclusive-jet cross sections measured by the ZEUS collaboration for for $Q^2 < 1$~GeV$^2$ \cite{ZEUS-prel-10-003}.
Jets were identified in the laboratory frame using the $k_T$-algorithm in the longitudinally inclusive mode  with $E_{t\;jet} > 17$~GeV and $-1 < \eta_{jet} < 2.5$. 
The measured cross section as a function of $\eta_{jet}$ is shown in Fig.~\ref{Fig:ZEUS} and compared to predictions from the \PYTHIA\  generator for different cuts of $p_{t\;min}$. It can be seen (Fig.~\ref{Fig:ZEUS} left) that a significant contribution from multi-parton interactions as implemented in \PYTHIA is needed, especially in the region  towards the outgoing proton remnant (large $\eta$). For higher $E_T$ values of the jet, the contribution from multiparton interactions is smaller, as can be seen from Fig.~\ref{Fig:ZEUS} (right), since the NLO calculation is within the experimental and theoretical uncertainties.
\subsection{Measurement of the underlying event}
\begin{wrapfigure}{r}{0.3\textwidth}
\centerline
{\includegraphics[width=0.25\textwidth]{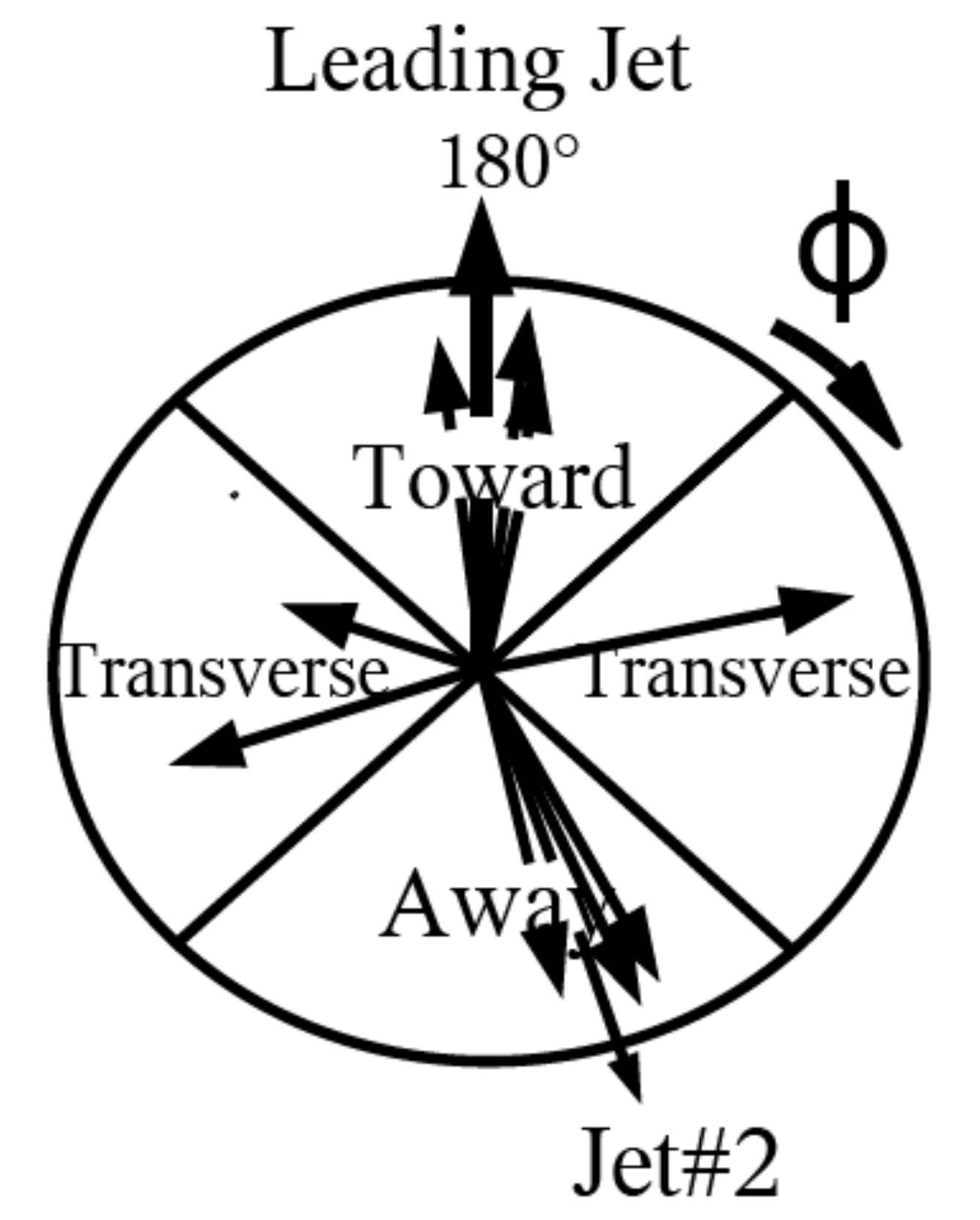}}
\caption{The definition of the various regions in azimuthal angle.}\label{Fig:Transregdijet}
\end{wrapfigure}

The H1 analysis~\cite{H1prelim-08-036} of the underlying event follows the one of CDF~\cite{Affolder:2001xt}. The jet with
the highest transverse momentum, (using the $k_T$-algorithm in the hadronic center of mass frame)  defines the azimuthal angle $\Phi^{*}$=0 (Fig.~\ref{Fig:Transregdijet}).
The region $|\Phi^{*}| < 60^o$ is defined as the
\textit{toward region}, and is expected to
contain all particles
belonging to the leading jet. The jet with the second highest transverse
momentum is often inside the \textit{away
region}, which is defined by  $|\Phi^{*}| > 120^o$. The
\textit{away region} is expected to host most of  the particles which
are needed to provide momentum balance in the event. The fact that the
jets are not required to be back-to-back implies that the phase space
is increased for production of additional jets. The \textit{transverse
regions},  $60^o < |\Phi^{*}| < 120^o$ are those where contributions
from the primary collision  should be relatively small and thus  effects from
underlying events should be most visible. Event by event a
\textit{high activity}- and \textit{low activity region} is defined,
depending on which region contains the highest scalar sum of the 
transverse momenta of charged particles. 

The observable $x_{\gamma}$, defined as
the fractional longitudinal momentum of the photon carried by the
parton involved  in the hard scattering, gives a handle to favour contributions
from resolved or direct photon processes.  In resolved
processes ($x_{\gamma} < 1$) the photon fluctuates into a quark pair and thus only a fraction of the photon momentum is entering the hard scattering. 

The transverse momenta of each of the two hardest jets in the event are required to have 
$E_{T\;jet} > 5$~GeV in $-1.5 <  \eta^{lab} <  1.5$.
The data are corrected for trigger inefficiencies, limited detector acceptance and
resolution. 

\begin{figure}[htb]
\centerline{\includegraphics[width=1\textwidth]{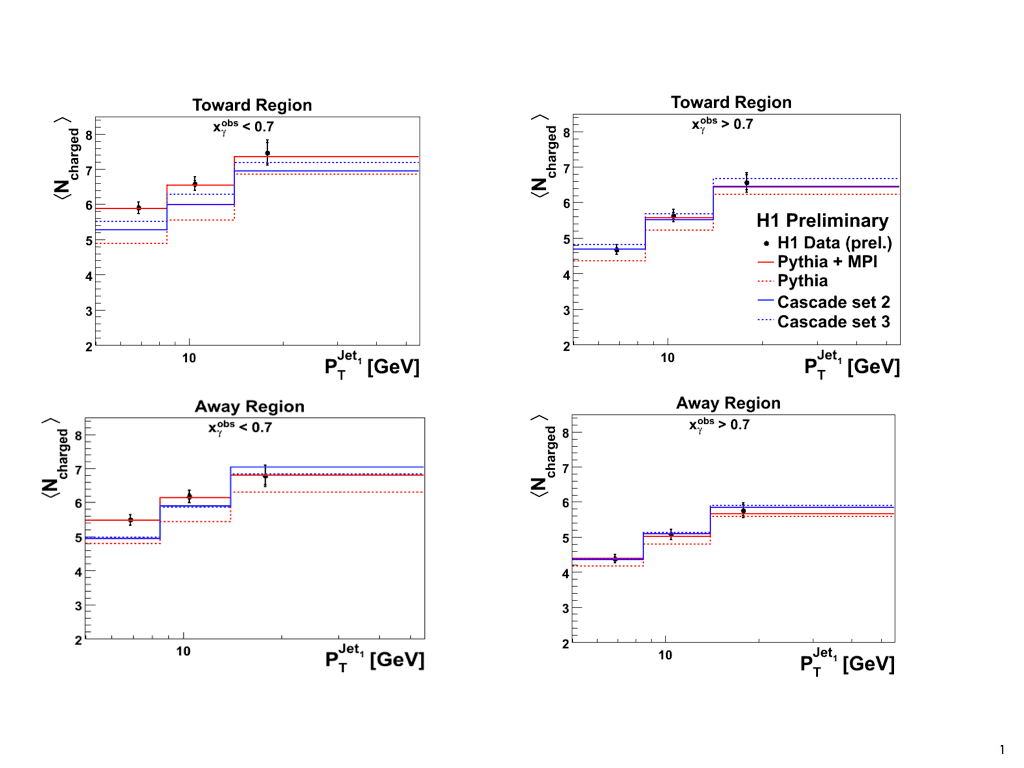}}
\caption{Average charged particle multiplicity in photoproduction as a function
of the transverse momentum of the leading jet, $E_{T\,jet1}$, for the
toward and away regions (upper and lower plots, respectively) and for
the low and high $x_\gamma^{obs}$ regions (left and right, respectively). The data
are compared to {\sc Pythia} with and without the MPI model as well as to the \CASCADE\ predictions with two different proton parton densities.}\label{Fig:Pyth62-HardScatt}
\end{figure}

\begin{figure}[htb]
\centerline{\includegraphics[width=1\textwidth]{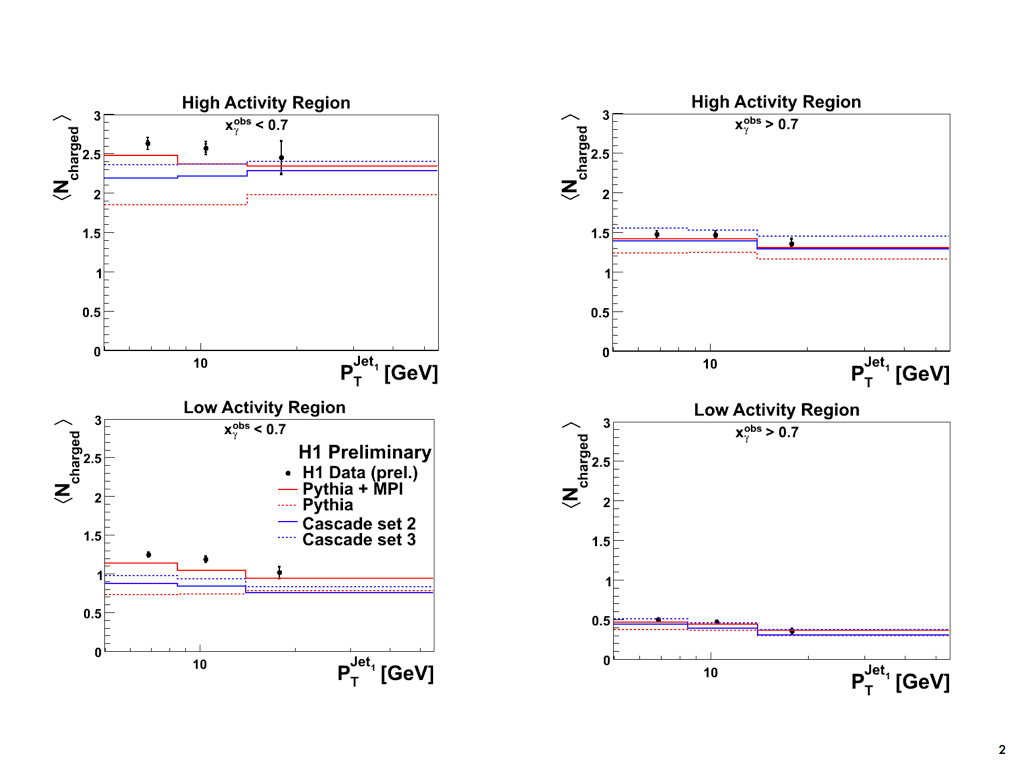}}
\caption{Average charged multiplicity in photoproduction as a function
of the transverse momentum of the leading jet, $E_{T\,jet1}$,
for the transverse high and low activity regions (upper and lower plots,
respectively) for the low and high $x_\gamma^{obs}$ regions (left and right,
respectively). The data are compared to {\sc Pythia} with and without MPI as well as to the \CASCADE\ predictions with two different proton parton densities.}\label{Fig:Pyth62-Trans}
\end{figure}

The average charged particle multiplicity, $\langle N_{charged}
\rangle$, for dijet events in photoproduction is shown in
Figs.~\ref{Fig:Pyth62-HardScatt} and \ref{Fig:Pyth62-Trans} as a function
of the transverse momentum of the leading jet, $E_{T\,jet1}$,
for the different azimuthal regions. The measurement is separated into
a resolved and a direct photon enriched region by applying the cuts
$x^{\textit{obs}}_{\gamma} < 0.7$ and $x^{\textit{obs}}_{\gamma} >
0.7$, respectively. 

The average particle multiplicities, $\langle N_{charged} \rangle$,
for the data range between 4-8 particles in the toward and away
region (see Fig.~\ref{Fig:Pyth62-HardScatt}), and between 0.5-2.5 particles in the transverse regions (see Fig.~\ref{Fig:Pyth62-Trans}. In
the toward and away region $\langle N_{charged} \rangle$ increases
with the transverse momentum of the leading jet, $E_{T\,jet1}$. In the transverse regions the
behaviour is opposite but much weaker with a decrease of the average multiplicity from the lowest to the highest $E_{T\,jet1}$ bin. The particle multiplicity is slightly higher in the toward region compared to the away region, and  significantly higher in the high activity region than
in the low activity region. 
In Fig.~\ref{Fig:Pyth62-HardScatt}  the data in the toward and away regions are compared to different MC
predictions. The charged particle multiplicities, which in these
regions are expected to be dominated by the hard process, are very well
described by \PYTHIA\, if contributions from MPI are included.  \CASCADE\,
which does not simulate any multiparton interactions, but is based on a
different evolution scheme for the parton showers, also describes the
data fairly well, although somewhat worse in the low $x^{\textit{obs}}_{\gamma}$ region. Please note, that \CASCADE\ has no explicit resolved photon contribution included.

The data in the low and high activity transverse regions are
shown in Fig.~\ref{Fig:Pyth62-Trans}. The models are somewhat underestimating the data in the resolved enriched region. Like in the toward and away regions, we see that also here the contribution from MPI in {\sc Pythia} gives a significantly better
description of the data especially for the small $x_{\gamma}$ in the high activity region(Fig.~\ref{Fig:Pyth62-Trans}). {\sc Cascade}
describes the data in the high $x^{\textit{obs}}_{\gamma}$ region,
but predicts a too low charged particle multiplicity in the low $x^{\textit{obs}}_{\gamma}$ region.
\section{Conclusion}
Measurements of jet production in the photoproduction region at HERA show that a significant part of the the cross section is not well described by leading order Monte Carlo generators in collinear factorisation. NLO calculations are in agreement with the measurement within the quoted experimental uncertainty at large $E_{T\; jet}$.
A dedicated measurement of the average charged particle multiplicity in jet events in photoproduction shows that in the transverse region (compared to the jet direction) the contribution from the underlying event becomes significant. Different approaches for the parton showering to simulate higher order corrections differ significantly. Thus the amount of multiparton interaction needed to described the measurements depends very strongly on the underlying approach to simulate higher order parton radiation. 

\section*{Acknowledgements}
Many thanks go to the organisers of this very interesting and very well organised workshop. I am also grateful to my colleagues in H1 and ZEUS for help in preparing the talk: Ll. Marti, K. Kr\"uger, C. Glasman and A. Levy.


\begin{footnotesize}

\end{footnotesize}


\end{document}